\documentclass[%
 reprint,
superscriptaddress,
 amsmath,amssymb,
 aps,
]{revtex4-1}

\usepackage{graphicx}
\usepackage{dcolumn}
\usepackage{bm}

\begin{document}

\preprint{APS/123-QED}

\title{Ground-state thermodynamic quantities of homogeneous spin-$1/2$ fermions from the BCS region to the unitarity limit}

\author{Munekazu Horikoshi}
\affiliation{Institute for Photon Science and Technology, Graduate School of Science, The University of Tokyo, 7-3-1 Hongo, Bunkyo-ku, Tokyo 113-8656, Japan}
\affiliation{Photon Science Center, Graduate School of Engineering, The University of Tokyo, 2-11-16 Yayoi, Bunkyo-ku, Tokyo 113-8656, Japan}

\author{Masato Koashi}
\affiliation{Institute for Photon Science and Technology, Graduate School of Science, The University of Tokyo, 7-3-1 Hongo, Bunkyo-ku, Tokyo 113-8656, Japan}
\affiliation{Photon Science Center, Graduate School of Engineering, The University of Tokyo, 2-11-16 Yayoi, Bunkyo-ku, Tokyo 113-8656, Japan}
\affiliation{Department of Applied Physics, Graduate School of Engineering, The University of Tokyo, 7-3-1 Hongo, Bunkyo-ku, Tokyo 113-8656, Japan}

\author{Hiroyuki Tajima}
\affiliation{Department of Physics, Keio University, 3-14-1 Hiyoshi, Kohoku-ku, Yokohama 223-8522, Japan}

\author{Yoji Ohashi}
\affiliation{Department of Physics, Keio University, 3-14-1 Hiyoshi, Kohoku-ku, Yokohama 223-8522, Japan}

\author{Makoto Kuwata-Gonokami}
\affiliation{Institute for Photon Science and Technology, Graduate School of Science, The University of Tokyo, 7-3-1 Hongo, Bunkyo-ku, Tokyo 113-8656, Japan}
\affiliation{Photon Science Center, Graduate School of Engineering, The University of Tokyo, 2-11-16 Yayoi, Bunkyo-ku, Tokyo 113-8656, Japan}
\affiliation{Department of Physics, Graduate School of Science, The University of Tokyo, 7-3-1 Hongo, Bunkyo-ku, Tokyo 113-0033, Japan}

\date{\today}

\begin{abstract}

We experimentally determined various thermodynamic quantities of interacting two-component fermions at the zero-temperature limit from the Bardeen-Cooper-Schrieffer (BCS) region to the unitarity limit. The obtained results are very accurate in the sense that the systematic error is within 4\% around the unitarity limit. Using this advantage, we can compare our data with various many-body theories. We found that an extended {\it T}-matrix approximation, which is a strong-coupling theory involving fluctuations in the Cooper channel, well reproduces our experimental results. We also found that the superfluid order parameter ${\it \Delta}$ calculated by solving the ordinary BCS gap equation with the chemical potential of interacting fermions is close to the binding energy of the paired fermions directly observed in a spectroscopic experiment and that obtained using a quantum Monte Carlo method. Since understanding the strong-coupling properties of a superfluid Fermi gas in the BCS-BEC (Bose-Einstein condensation) crossover region is a crucial issue in condensed matter physics and nuclear physics, the results of the present study are expected to be useful in the further development of these fields.

\begin{description}
\item[PACS numbers]
\end{description}
\end{abstract}

\pacs{Valid PACS appear here}
\maketitle


\section{\label{sec1}Introduction}

A many-body system of fermions interacting with $s$-wave scattering length $a$ is a fundamental model that extends the ideal Fermi gas model for various interacting Fermi systems. Understanding the ground state properties of fermions in the unitary regime, where the absolute value of the scattering length is larger than the inter-particle distance $d \sim k_F^{-1}$, is crucial in condensed matter physics and nuclear physics, where $k_F$ is the Fermi wave number. For example, the relation between number density $n$ and the internal energy density $\mathcal{E}$ with neutron-neutron scattering length $a_{\rm NN}=-18.63$~fm \cite{PhysRevC.77.054002} gives the equation of state (EOS) for dilute neutron matter. The EOS characterizes the symmetry energy of nucleons, the neutron skin for neutron-rich nuclei, and the inner structure of neutron stars \cite{doi:10.1146/annurev-nucl-102711-095018}. Since neutrons have a negative scattering length, dilute neutron matter exists in the interaction range from the Bardeen-Cooper-Schrieffer (BCS) limit to the unitarity limit at different neutron densities. Therefore, many-body physics from the BCS limit to the unitarity limit is common to both condensed matter physics \cite{PhysRev.186.456} and nuclear physics \cite{doi:10.1146/annurev-nucl-102014-021957}.

Ultracold atomic gases provide an ideal research environment in which we can investigate such many-body Fermi systems universally and systematically \cite{zwerger2011bcs,RevModPhys.80.1215,RevModPhys.80.885}. Many-body systems composed of ultracold atoms have an ideal hierarchy of length and energy, where the inter-particle distance and the wavelength of matter are sufficiently long compared to the size of the short-range interaction potential. Furthermore, it is possible to tune the scattering length between two fermions using Feshbach resonances \cite{RevModPhys.82.1225}. These features realize universal many-body systems, where various physical phenomena are independent of  details of particles \cite{PhysRevLett.92.090402}.

The ground state of many-body fermions interacting with an $s$-wave scattering length is an $s$-wave superfluid state, which has a nonzero superfluid order parameter ${\it \Delta}$ of the paired fermions and fluctuations of the order parameter. The origin of such superfluid fluctuations is the repetition of pair formations and their dissociations as well as non-condensed pairs that are kicked out from the condensate. Far below the superfluid phase transition temperature $T_c$, where thermal excitations are almost absent, many-body corrections to physical quantities are dominated by quantum fluctuations associated with superfluid fluctuations. 

At the BCS limit, the BCS mean-field (MF) wavefunction has been considered to be an adequate approximation to describe the ground state properties because the influence of quantum fluctuations appears to be small due to weak interaction between fermions. In this BCS-MF approximation, the magnitude of the order parameter ${\it \Delta}$ is equivalent to the binding energy of a paired fermion and to the energy gap in single-particle excitations \cite{leggett2006quantum}. The relation between the thermodynamic quantities and ${\it \Delta}$ is given by the gap equation and the number equation. The influence of quantum fluctuations on the ground state energy has also been theoretically analyzed at the BCS limit. The thermodynamic quantities of such fermions have been expressed analytically up to the order of $(k_Fa)^2$ beyond the MF approximation \cite{PhysRevA.77.023626}.

In the unitary regime, on the other hand, it is not obvious whether the BCS-MF wave function is an adequate approximation for describing the ground state properties because considerable quantum fluctuations are induced by strong interaction between fermions. There are no analytical formulas to give the order parameter, the binding energy of a paired fermion, the thermodynamic quantities, or the relations among them. Thus far, the binding energy of a paired fermion \cite{PhysRevLett.101.140403}, a single-particle excitation spectrum \cite{PhysRevLett.109.220402,1742-6596-467-1-012010,PhysRevLett.114.075301}, and internal energy density $\mathcal{E}$ \cite{Navon729}, which is the ground state energy per unit volume, have been experimentally determined in the unitary regime. Other thermodynamic quantities, such as pressure \cite{Navon729,Ku563}, isothermal compressibility \cite{Ku563, PhysRevLett.106.010402}, speed of sound \cite{ PhysRevLett.98.170401}, and contact density \cite{PhysRevLett.114.075301} were also measured. However, the magnitude of the order parameter and chemical potential $\mu$ have not yet been determined experimentally in the unitary regime. While several theories have demonstrated qualitative agreement with these experimental results, quantitative evaluation of these theories has not been achieved because of experimental accuracies, lack of data points, and inhomogeneity effects.

In the present study, we determine the relations among thermodynamic quantities, such as pressure $P$, number density $n$, internal energy density $\mathcal{E}$, chemical potential $\mu$, isothermal compressibility $\kappa$, and contact density $\mathcal{C}$, for homogeneous spin-$1/2$ superfluid fermions at the zero-temperature limit from the BCS region to the unitarity limit. All of these quantities were determined within 4\% systematic error around the unitarity limit without using any model functions. We provide their thermodynamic functions in universal form as a function of an interaction parameter. Therefore, these functions are applicable to other similar many-body Fermi systems, such as dilute neutron matter. We compared our experimental data with various many-body theories and found that one theoretical model can reproduce our data. We also calculated the superfluid order parameter by solving the ordinary BCS gap equation with the determined chemical potential. We compare the obtained results with the binding energies of paired fermions, which were determined in previous studies through a spectroscopic experiment and quantum Monte Carlo calculation.

The remainder of the present paper is organized as follows. In Section \ref{sec2}, we introduce the theoretical framework used in our data analysis. In Section \ref{sec3}, we explain our experimental procedure, data analysis, and experimental results. In Section \ref{sec:Result}, we compare our experimental results with the results of various previous experimental and theoretical studies and discuss calculation of the superfluid order parameter. Finally, in Section \ref{sec5}, we conclude this study.

\section{\label{sec2}Theoretical framework}

\subsection{\label{sec:thermodynamics}Thermodynamics of homogeneous fermions from the BCS limit to the unitarity limit at zero temperature}

We consider homogeneous spin-$1/2$ fermions at $T=0$, which have a number density of $n_\uparrow$~=~$n_\downarrow$~=~$n/2$ and a chemical potential of $\mu_\uparrow$~=~$\mu_\downarrow$~=~$\mu$ for each spin state. We assume that the interaction between fermions in different spin states is modeled by the $s$-wave scattering length $a$. In the following, for convenience, we use $a^{-1}$ instead of $a$. We consider the parameter region $a^{-1}\leq0$, where the Fermi system ranges from the BCS limit ($a^{-1}=-\infty$) to the unitarity limit ($a^{-1}=0$). Let $m$ be the mass of the fermions, and let $\hbar$ be the reduced Plank constant.

The internal energy density $\mathcal{E}(n,a^{-1})$ is a function of $n$ and $a^{-1}$.
Let us write its differential form as
\begin{equation}
d\mathcal{E}=\mu dn-\left( \frac{\hbar^2}{4\pi m}\mathcal{C} \right)da^{-1}
\label{eq3}.
\end{equation}
The thermodynamic quantity $\mathcal{C}$ defined in the above relation is called the contact density \cite{Tan20082952,Tan20082971,Tan20082987,PhysRevA.78.053606,PhysRevA.79.023601,PhysRevA.86.013626}.
The pressure $P$ is derived from $\mathcal{E}(n,a^{-1})$ as
\begin{equation}
P=\mu n-\mathcal{E}
\label{eq4},
\end{equation}
The functional dependence $P(\mu, a^{-1})$ has a simple differential form:
\begin{equation}
dP=nd\mu+\left( \frac{\hbar^2}{4\pi m}\mathcal{C} \right)da^{-1}
\label{eq5}.
\end{equation}
Isothermal compressibility $\kappa$ and speed of sound $v$ have the same thermodynamic relations as an ideal Fermi gas under a fixed scattering length:
\begin{gather}
\kappa=\frac{1}{n}\left( \frac{\partial n}{\partial P} \right)_{a^{-1}}=\frac{1}{n^2}\left( \frac{dn}{d\mu} \right)_{a^{-1}} ,\label{eq8}\\
v=\sqrt{\frac{1}{mn\kappa}}.\label{eq9}
\end{gather}

\subsection{\label{sec:Dimensionless}Dimensionless thermodynamic functions}

Here, we use the fact that the theory involves only two constants, $m$ and $\hbar$. If we change the unit of length by a factor of $\lambda$ and that of time by $\lambda^2$, the values of the two constants remain unchanged. Hence, the functional dependence of any dimensionless qualities $A$ on $n$ and $a^{-1}$ should satisfy $A(n, a^{-1})=A(\lambda^3 n, \lambda a^{-1})$. This implies that $A$ can be written as a function $A(x)$ of a parameter $x$ proportional to $a^{-1}n^{-1/3}$ \cite{PhysRevLett.92.090402}. Similarly, when a dimensionless quantity $B$ is a function of $\mu$ and $a^{-1}$, it should be written as a function $B(X)$ of a parameter $X$ proportional to $a^{-1}\mu^{-1/2}$. 

In the present paper, we choose the two parameters to be
\begin{align}
\text{grand-canonical interaction parameter:}\nonumber\\
X(\mu,a^{-1})=\frac{1}{k_{\mu}(\mu)a},\label{eq13}\\
\text{canonical interaction parameter:}\nonumber\\
x(n,a^{-1})=\frac{1}{k_{F}(n)a}.\label{eq14}
\end{align}
Here, we define two wave numbers as $k_{\mu}(\mu)=\sqrt{2m\mu}/\hbar$ and  $k_F(n)=\sqrt{2m\varepsilon_F (n)}/\hbar$, where $\varepsilon _F(n)=\frac{\hbar^2}{2m}(3\pi^2n)^{2/3}$ is the Fermi energy for ideal fermions.
For each thermodynamic quantity, we define an associated dimensionless thermodynamic function by normalizing it with the value for an ideal Fermi gas as follows:
\begin{align}
\text{pressure:~}f_P(X)=\frac{P}{P_0(\mu)},\label{eq15}\\
\text{number density:~}f_n(X)=\frac{n}{n_0(\mu)},\label{eq16}\\
\text{internal energy density:~}f_\mathcal{E}(x)=\frac{\mathcal{E}}{\mathcal{E}_0(n)},\label{eq17}\\
\text{chemical potential:~}f_{\mu}(x)=\frac{\mu}{\varepsilon_F (n)},\label{eq18}\\
\text{isothermal compressibility:~}f_{\kappa}(x)=\frac{\kappa}{\kappa_0 (n)},\label{eq19}\\
\text{speed of sound:~}f_{v}(x)=\frac{v}{v_0 (n)},\label{eq20}
\end{align}
where $P_0 (\mu)=\frac{2}{15\pi^2} \left(\frac{2m}{\hbar^2}\right)^{3/2} \mu^{5/2}$, $n_0 (\mu)=\frac{5}{2} \frac{P_0 (\mu)}{\mu}$, $\mathcal{E}_0 (n)=\frac{3}{5} n\varepsilon_F (n)$, $\kappa_0 (n)=\frac{3}{2} \frac{1}{n\varepsilon_F (n)}$, and $v_0(n)=\left( mn\kappa_0(n) \right)^{-1/2}$.
An exception is the contact density $\mathcal{C}$, which vanishes for the ideal gas.
Here, we choose its dimensionless thermodynamic function as
\begin{equation}
\text{contact density:~}f_{\mathcal{C}}(x)=\frac{\mathcal{C}}{k_F (n)^4}.
\label{eq21}
\end{equation}

In an experiment with interacting Fermi gas in a trap potential in Section \ref{sec3}, we obtain data including various combinations of $(\mu, P)$ with a fixed value of the scattering length $a$.
It is convenient for data analysis to introduce dimensionless versions of $\mu$ and $P$ by using normalizing factors solely determined from $a$ as
\begin{equation}
\begin{cases}
\; \mathcal{G}=\frac{\mu}{\varepsilon_a}, \\
\; \chi=P\frac{a^3}{\varepsilon_a},
\end{cases}
\label{relationGchi1}
\end{equation}
where we define energy as $\varepsilon_a=\frac{\hbar^2}{2ma^2}$.
According to the dimensionless analysis discussed above, we can see that $\mathcal{G}$ is given as a function of $\chi$, i.e., $\mathcal{G}(\chi)$.
Since $X$ and $f_P$ are simply related to $\mathcal{G}(\chi)$ as
\begin{equation}
\begin{cases}
\; X=-\mathcal{G}(\chi)^{-1/2}, \\
\; f_P=-\frac{15\pi^2}{2}\chi \mathcal{G}(\chi)^{-5/2},
\end{cases}
\label{relationGchi}
\end{equation}
$\mathcal{G}(\chi)$ can be directly converted to $f_P(X)$.

The function $f_n(X)$ is derived from $f_P(X)$ as
\begin{equation}
f_n(X)=f_P(X)-\frac{X}{5}\frac{df_P(X)}{dX}.
\label{relationnP}
\end{equation}
The function $f_\mathcal{E}(x)$ is constructed by
\begin{equation}
\begin{cases}
\; f_\mathcal{E}=\frac{5f_n(X)-2f_P(X)}{3f_n^{5/3}(X)}, \\
\; x=Xf_n (X)^{-\frac{1}{3}}.
\end{cases}
\label{relationB}
\end{equation}
The function $f_\mu(x)$ is derived from $f_\mathcal{E}(x)$ as
\begin{equation}
f_\mu(x)=f_\mathcal{E}(x)-\frac{x}{5}\frac{df_\mathcal{E}(x)}{dx}.
\label{relationmuE}
\end{equation}
When $f_\mu(x)$ is given, $f_\kappa(x)$ can be determined by
\begin{equation}
f_\kappa(x)=\left \{ f_\mu(x)-\frac{x}{2}\frac{df_\mu(x)}{dx} \right \}^{-1}.
\label{relationmuk}
\end{equation}
The function $f_v(x)$ has a simple relation to $f_\kappa(x)$,
\begin{equation}
f_v(x)=f_\kappa(x)^{-1/2}.
\label{relationvk}
\end{equation}
The function $f_\mathcal{C}(x)$ is obtained from the derivative of $f_\mathcal{E}(x)$ as,
\begin{equation}
f_\mathcal{C}(x)=-\frac{2}{5\pi}\frac{df_\mathcal{E}(x)}{dx}.
\label{relationCE}
\end{equation}

Equations~(\ref{relationnP}) through (\ref{relationmuk}) indicate that there is a universal value $\xi$ at the unitarity limit ($X=x=0$), which is $\xi=f_P (0)^{-2/3}=f_n (0)^{-2/3}=f_\mathcal{E} (0)=f_\mu (0)=f_\kappa (0)^{-1}$.
Note that $\xi$ is sometimes called the Bertsch parameter \cite{zwerger2011bcs}.

\subsection{Behavior of the dimensionless functions at the BCS limit}

The asymptotic behavior of $f_\mathcal{E} (x)$ at the BCS limit is known up to the second order of $1/x$ \cite{fetter2003quantum, PhysRevA.77.023626} as
\begin{equation}
f_\mathcal{E}^{\rm Asym}(x\rightarrow -\infty)=1+\frac{10}{9\pi}x^{-1}+\frac{4(11-2{\rm log}2)}{21\pi^2}x^{-2}. \label{eq34}
\end{equation}
Here, we omit the condensation energy term \cite{PhysRevA.77.023626}.
Moreover, $f_\mu^{\rm Asym}(x)$ is given by Eq.~(\ref{relationmuE}) with $f_\mathcal{E}^{\rm Asym}(x)$, and $f_P^{\rm Asym}(X)$ is obtained using the following relations: 
\begin{equation}
\begin{cases}
\; f_P=\frac{5f_\mu(x)-3f_\mathcal{E}(x)}{2f_\mu^{5/2}(x)}, \\
\; X=xf_\mu (x)^{-\frac{1}{2}}. \\
\end{cases}
\label{relationD}
\end{equation}
The asymptotic behavior of the other thermodynamic quantities at the BCS limit can be obtained from $f_P^{\rm Asym}(X)$ using Eqs.~(\ref{relationnP}) through (\ref{relationCE}).

\section{\label{sec3}Experiment}

\subsection{\label{sec:Acquisition}Data acquisition}

\begin{figure}[!tb]
\includegraphics{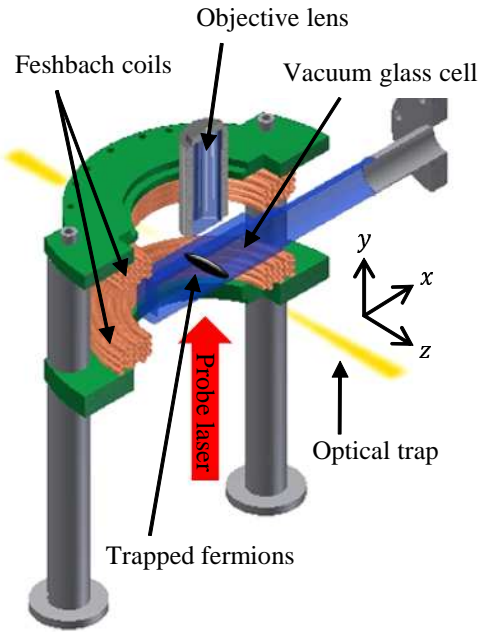}
\caption{\label{fig0}
Schematic diagram of the experimental apparatus.
The $z$ axis is defined as the axial direction of the optical trap.
}
\end{figure}

We have an experimental apparatus that can produce dual Bose-Einstein condensates (BECs) of paired $^6$Li (fermion) and one spin state of $^7$Li (boson) in the hybrid trap of an optical dipole trap (ODT) and a magnetic trap (MT) \cite{HoriSuperfluid}. A schematic drawing of the experimental apparatus is shown in Figure~\ref{fig0}. Detailed information about our experimental apparatus and experimental procedure is provided in a previous study \cite{HoriSuperfluid}. In this experiment, only $^6$Li atoms were used.
The shape of the trapping potential $U_{\rm trap}$ is analytically given without using a harmonic approximation in terms of the parameters of the laser power, the wavelength, the beam waists of the ODT, and the magnetic curvature of the MT. The trap has an elliptic symmetry and can be written as $U_{\rm trap}(x, y, z)=U_{\rm trap}(\rho, z)$ with $\rho=\sqrt{x^2+\eta^2(z)y^2}$, where $\eta(z)$ is the ellipticity.

The ODT is produced by a focused 1070-nm laser beam. The $1/e^2$ beam radii are $(w_{0x},w_{0y}) = (43.5, 46.9)$~$\mu$m, and the final laser power after evaporative cooling is $P_{\rm ODT}$=45~mW. The depth of the ODT is 38~uK, and the trapping frequency realized by the ODT is $(\omega_x, \omega_y, \omega_z)$~=~$2\pi \times (250, 230, 1.3)$~Hz. The MT is produced by the magnetic curvature of a bias magnetic field for the Feshbach resonance and produces magnetic curvature $\omega_{\rm mag}=2\pi \times 0.24\sqrt{B}$~Hz in the $z$ direction, where $B$ is the produced bias magnetic field in Gauss. The effective trapping frequency along the $z$ direction is given by $\omega_{z,\rm{eff}}=\sqrt{\omega_z^2+\omega_{\rm mag}^2}$. Typically, $2\times 10^5$ degenerate fermions are prepared in this experiment.

$^6$Li atoms have a broad $s$-wave Feshbach resonance for collisions between the two lowest spin states: $\left| 1 \right \rangle$~$\equiv$~$\left| m_L= 0,~m_S=-1/2,~m_I=+1\right\rangle$ and $\left| 2 \right \rangle$~$\equiv$~$\left| m_L= 0,~m_S=-1/2,~m_I= 0\right\rangle$, where $m_L$, $m_S$, and $m_I$ are, respectively, the electronic orbital angular momentum projection, the electronic spin projection, and the nuclear spin projection. The scattering length is given by $a(B)=a_{\rm bg}\left( 1+\frac{W_{\rm res}}{B-B_{\rm res}} \right)$ with parameters of $a_{\rm bg}$=$-$1582$a_0$, $B_{\rm res}$=832.18 Gauss, and $W_{\rm res}$=262.3 Gauss, where $a_0$ is the Bohr radius \cite{PhysRevLett.110.135301}. Then, the Fermi system is in the BEC region ($a>0$) at $B<B_{\rm res}$. On the other hand, the Fermi system is in the BCS region ($a<0$) at $B>B_{\rm res}$. At $a(B=B_{\rm res})=\pm \infty$, the Fermi system reaches the unitarity limit.

In the present study, we are interested in the ground state properties of the interacting fermions from the BCS limit to the unitarity limit.
In order to prepare highly degenerate Fermi gas ($t=k_BT/\varepsilon _F(n) \ll1$) in the interaction region, we produced an almost pure molecular BEC consisting of two fermions in the state of $\left| 1 \right \rangle$ and $\left| 2 \right \rangle$ at 777~Gauss in the BEC regime. We then changed the scattering length adiabatically by sweeping the magnetic field to 832.18$\sim$1050~Gauss in 1.5~s.

\begin{figure}[!tb]
\includegraphics{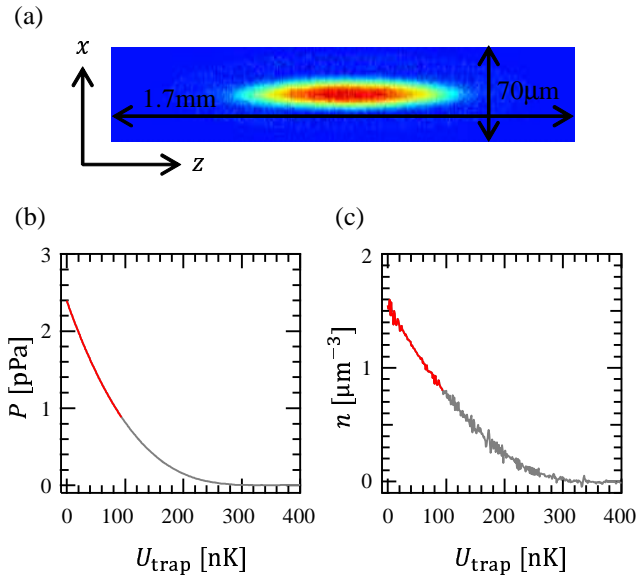}
\caption{\label{fig1}
Typical experimental data at 882.86~Gauss.
(a) Absorption image.
(b) Local pressure $P(U_{\rm trap})$.
(c) Local number density $n(U_{\rm trap})$.
The red curves in (b) and (c) show the data region used in the analysis, where the number density is larger than half of the peak density.
}
\end{figure}

We can accurately obtain their density distribution {\it in situ} using the imaging techniques developed in a previous study \cite{HoriAbsorption}.
We measured the optical depth (OD) of the trapped fermions at each magnetic field, as shown in Figure~\ref{fig1}a.
From the $OD$ and the absorption cross-section $\sigma_{\rm abs}$, we evaluated the column density $\bar{n}(x, z)=\int_{-\infty}^{+\infty}n(x, y, z)dy$ by $\bar{n}(x, z)=OD(x, z)/\sigma_{\rm abs}$.
We determined the effective value of $\sigma_{\rm abs}$ within 4\% uncertainty \cite{HoriAbsorption}.
Note that this uncertainty results in one of the systematic errors in our final results.

The local pressure of the trapped fermions can be calculated from the column density ${\bar n}(x,z)$ and the trapping potential $U_{\rm trap}(\rho, z)$ using the following formula \cite{ho2010obtaining,HoriAbsorption,Werner}:
\begin{multline}
P(\rho,z)=\frac{\eta(z)}{\pi}\int_{\rho}^{\infty}dx~{\bar n}(x,z)\left[ \frac{\frac{\partial U_{\rm trap}}{\partial \rho}(x,z)}{(x^2-\rho^2)^{1/2}} \right. \\
\left. + \int_{\rho}^x d\rho '~\frac{\rho '\frac{\partial U_{\rm trap}}{\partial \rho}(x,z)-x\frac{\partial U_{\rm trap}}{\partial \rho}(\rho ',z)}{(x^2-\rho '^2)^{3/2}} \right],
\label{eqLocalPressure}
\end{multline}
which is derived from the Gibbs-Duhem equation [Eq.~(\ref{eq5})], the local density approximation (LDA), i.e., $\mu=\mu_0-U_{\rm trap}$, and the inverse Abel transformation of the column density, i.e., $n(\rho,z)=-\frac{\eta(z)}{\pi }\int_{\rho}^{\infty}\frac{1}{\sqrt{x^2-\rho ^2}}\frac{\partial \bar{n}(x,z)}{\partial x}dx$.
This calculation can be carried out without the knowledge of the global chemical potential $\mu_0$ of the trapped system. Finally, we can obtain pressure $P$ as a function of the trapping potential $U_{\rm trap}$ by relating $P(\rho, z)$ and $U_{\rm trap}(\rho, z)$ at each position.

Figure~\ref{fig1}b shows a typical example of $P(U_{\rm trap})$ at 882.86~Gauss.
The data are averaged at 241 potential heights in the region of $U_{\rm trap}/k_B \in [0, 600~{\rm nK}]$ with an interval of $\Delta U_{\rm trap}/k_B=2.5~{\rm nK}$.
We also evaluated the local number density from the local pressure $P(U_{\rm trap})$ according to $n(U_{\rm trap})=-\frac{dP(U_{\rm trap})}{dU_{\rm trap}}$ under the LDA, as shown in Figure~\ref{fig1}c.

We extracted $P(U_{\rm trap})$ and $n(U_{\rm trap})$ in the highly degenerate region, where the number density is larger than half of the peak density for the following data analysis, because $t=k_BT/\varepsilon _F(n) \propto n^{-2/3}$.
The red curves plotted in Figures \ref{fig1}b and \ref{fig1}c show the regions.
We acquired data sets of $\{P(U_{\rm trap}), n(U_{\rm trap}), a(B)^{-1}\}$ at 44 magnetic fields of from 832.54 to 1,050 Gauss for the BCS region and of 832.18 Gauss for the unitarity limit.
The 44 magnetic fields were chosen in order that the canonical interaction parameter $ x\left(n(U_{\rm trap}), a(B)^{-1} \right)$ obtained at $B$ overlaps one another. In this experiment, we repeated data acquisition eight times at each magnetic field to decrease statistical errors.

We hereinafter express physical quantities at $B$ by adding superscript $B$, for example, $P^{B}$, $n^{B}$, and $U_{\rm trap}^B$.

\subsection{\label{sec:Compressibility}Evaluation of $f_\kappa(x)$}

\begin{figure}[tb!]
\includegraphics{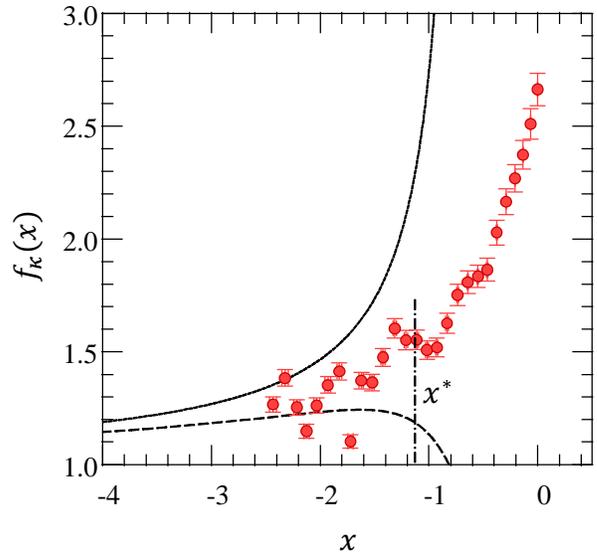}
\caption{\label{FigCompressibilityResult}
Experimental data of isothermal compressibility from the BCS region to the unitarity limit. The red circles are experimental data. The dotted curve and the dashed curve show their asymptotic behavior at the BCS limit up to the first and second order of $1/x$, respectively. The vertical dash-dot line indicates interaction parameter $x^*$, where superfluid transition occurs. The fermions are in the superfluid state in the region of $x>x^*$.
}
\end{figure}

The dimensionless isothermal compressibility $f_{\kappa}(x)$ can be constructed model-independently from data sets of $\{P^B(U_{\rm trap}^B), n^B(U_{\rm trap}^B)\}$.
We calculated $f_{\kappa}^B(U_{\rm trap}^B)$ and $x^B(U_{\rm trap}^B,B)$ from these sets using the thermodynamic relation of Eq.~(\ref{eq8}) and the definitions of $x(n,a^{-1})$ and $\kappa_0(n)$.
We collected $f_{\kappa}^B(x^B)$ from the BCS region to the unitarity limit and averaged the values at the given $x$, as shown by the red circles in Figure~\ref{FigCompressibilityResult}.
The error bars indicate the systematic errors in $f_\kappa$ and $x$ caused by the uncertainty of the absorption cross-section $\sigma_{\rm abs}$.
From the error propagation rule, these errors are $\frac{\delta f_\kappa}{f_\kappa}=\frac{2}{3}\frac{\delta \sigma_{\rm abs}}{\sigma_{\rm abs}}=2.7\%$ and $\frac{\sigma x}{x}=\frac{1}{3}\frac{\delta \sigma_{\rm abs}}{\sigma_{\rm abs}}=1.3\%$.
The statistical errors are within the error bars.

At the unitarity limit, our data indicate that $f_\kappa(0)=2.66(7)$, which corresponds to the universal value of $\xi=0.375(10)$. This value is very consistent with the accurate experimental value of 0.376(4) determined by measuring thermodynamics for homogeneous unitary Fermi gases \cite{Ku563}. In the BCS region, our data approach the theoretical asymptotic behavior of $f_\kappa(x)$.

Experimentally, it is impossible to prepare fermions at zero temperature, even if we start from a molecular BEC.
Therefore, experimental temperature $T_{\rm exp}$ always has a nonzero value, i.e., $t_{\rm exp}>0$.
Since the superfluid critical temperature $t_c$ decreases monotonically to zero from the unitarity limit to the BCS limit as a function of the interaction parameter $x$, it is inevitable that $t_c (x)$ intersects $t_{\rm exp}$ somewhere between the two interaction limits.
Here, we define the transition point as $t_{\rm exp}=t_c (x^* )$.

Based on another experiment, as described in Appendix~\ref{sec:Temperature}, the transition point was determined to be $x^*=-1.14$, and the temperature parameter of the fermions was estimated to be $0.06\lesssim t_{\rm exp}\lesssim 0.1$ for our experimental condition.
We indicate the superfluid transition point $x^*$ with a vertical dash-dot line in Figure \ref{FigCompressibilityResult}.
In the region of $x>x^*$, the fermions are in the superfluid state.
A cusp in $f_\kappa(x)$ appears around the transition point, as shown in Figure \ref{FigCompressibilityResult}, which is similar to that observed in a previous study \cite{Ku563}.  

According to the previous experiment \cite{Ku563} and a theoretical calculation \cite{PhysRevA.78.023625}, our temperature range of $0.06\lesssim t_{\rm exp}\lesssim 0.1$ is low enough that we can consider our Fermi system as the ground state at the unitarity regime, because a large amount of fermions are in the superfluid state, which has zero entropy.

\subsection{\label{sec:Construction}Construction of $\mathcal{G}(\chi)$}

\begin{figure*}[tb!]
\includegraphics{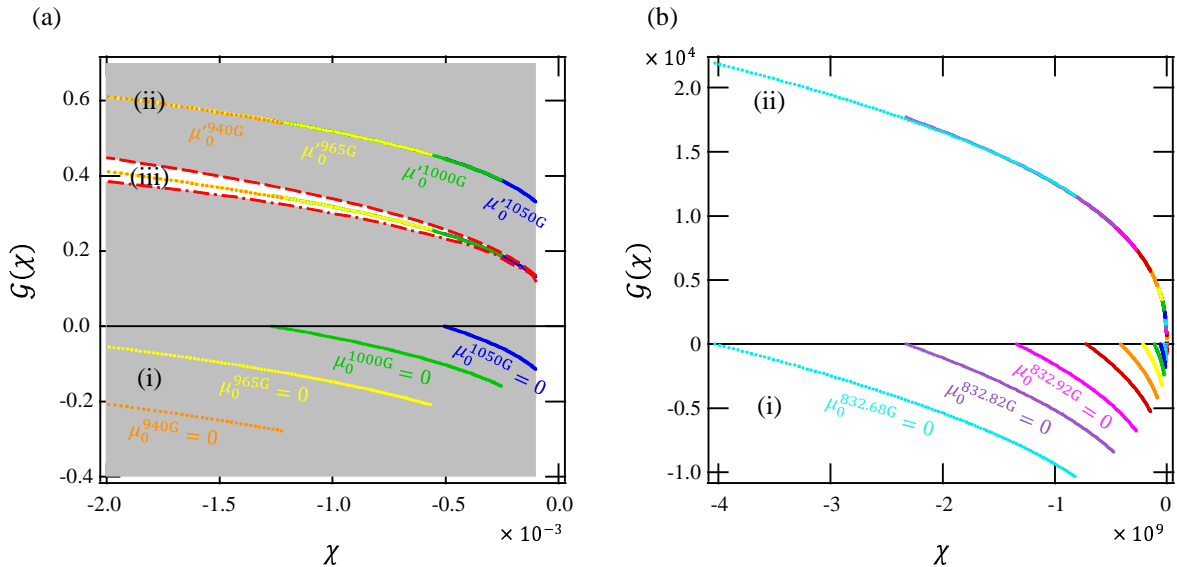}
\caption{\label{fig3}
Construction of dimensionless function $\mathcal{G}(\chi)$.
The different colors indicate the contributions of data taken at each Feshbach magnetic field.
(a) Around the BCS limit.
The dash-dot red curve and the dashed red curve indicate the lower and upper limits of $\mathcal{G}(\chi)$.
The gray areas correspond to the forbidden area for $\mathcal{G}(\chi)$.
(i) Unconnected data with $\mu_0^B=0$ for all data.
(ii) Connected data located in the forbidden area.
(iii) Connected data for which the offset value is tuned appropriately.
(b) Around the unitarity limit.
(i) Unconnected data with $\mu_0^B=0$ for all data.
(ii) Connected data for which the offset value is tuned appropriately.
}
\end{figure*}

In order to determine the other thermodynamic quantities, it is necessary to determine the chemical potential $\mu$ or contact density $\mathcal{C}$ along with $P$ and $n$.
In this experiment, we apply an almost homogeneous magnetic field to the gas, which implies that the Fermi gas has a homogeneous scattering length. In this case, it is impossible to determine $\mathcal{C}$ from the gradient of $P$ relative to $a^{-1}$ under the same $\mu$ for each data.

The local chemical potential $\mu^B$ is given by $\mu^B=\mu_0^B-U_{\rm trap}^B$ under the LDA, where $\mu_0^B$ is the global chemical potential at magnetic field $B$.
If we have experimental data corresponding to the fermions in the ground state at the BCS limit, we can determine $\mu_0^B$ in such a way that $f_P(X)$ matches the asymptotic theory at the BCS limit.
In this case, $\mu_0^B$ toward the unitarity limit can be determined by iterative fitting from the BCS limit with the same principle as demonstrated in a previous study \cite{1367-2630-12-10-103026}.
However, as shown in Figure \ref{FigCompressibilityResult}, it is difficult to judge whether our data reaches the BCS limit, and the data contains some finite temperature effects at $x<x^*$ because the fermions are in the normal state there.
Consequently, this method cannot be used to determine $\mu^B$ for our data.

Although the value of $\mu_0^B$ for each value of $B$ is unknown, we can still determine the relative difference among the various values of $B$ using the general properties of dimensionless parameters described in Section \ref{sec:Dimensionless}.
Equation (\ref{relationGchi1}) and the LDA suggest that all of the experimental data acquired at various Feshbach magnetic fields, $B$, should be related through a common function $\mathcal{G}(\chi)$ as 
\begin{equation}
\mathcal{G}\left( \chi\left( U_{\rm trap}^B, B \right), B \right)=\frac{\mu_0^B-U_{\rm trap}^B}{\varepsilon_{a(B)}}
\label{dless5},
\end{equation}
where $\chi\left( U_{\rm trap}^B,B \right)=P^B(U_{\rm trap}^B)\frac{a(B)^3}{\varepsilon_{a(B)}}$.
We plot $(\chi,\mathcal{G})$ with different choices of $\mu_0^B$ for all data in Figure~\ref{fig3}.
In the case of $\mu_0^B=0$ for all $B$, the plots do not overlap each other as the curves (i).
However, we can overlap the plots model-independently by simply adjusting each value of $\mu_0^B$ from the BCS limit ($B=1,050$~G) to the unitarity limit, as in the curves (ii).
This leaves only a single ambiguous parameter, namely, the arbitrary choice of $\mu_0^{B=1050{\rm G}}$. As a result, the experimentally constructed value $\mathcal{G}(\chi)$ deviates from the true value $\mathcal{G}_{\rm true}(\chi)$ by a constant offset as $\mathcal{G}(\chi)=\mathcal{G}_{\rm true}(\chi)+\Delta \mathcal{G}$, where $\Delta \mathcal{G}$ is independent of $\chi$.

We can derive a lower and an upper bound on the offset value in the following way.
The contact density should be positive in the BCS-BEC crossover ($\mathcal{C} >0$), and $\mathcal{C}$ is related to $f_P(X)$ by $\frac{\mathcal{C}}{k_\mu^4}=\frac{4}{15\pi}f'_P(X)$, which can be derived from Eq.~(\ref{eq5}). Then, $f_P(X)$ must have positive gradient $f_P'(X)>0$ at an arbitrary interaction parameter $X$. This restriction gives the lower bound on $\mathcal{G}(\chi)$ through Eq.~(\ref{relationGchi}) as
\begin{equation}
\mathcal{G}(\chi)>\frac{5}{2}\chi \mathcal{G}'(\chi).
\label{condition1}
\end{equation}
The border of this condition is plotted by the dash-dot red curve in Figure \ref{fig3}a.
This condition should be satisfied for the entire experimental curve $\mathcal{G}(\chi)$.
The condition is most restrictive at $\chi_{\rm max}=-1.0\times10^{-4}$, which is the maximum value achieved in this experiment.
Thus, the lower bound is given by $\mathcal{G}(\chi_{\rm max})>0.13$, which gives a lower bound on the experimental range of $X$ from Eq.~(\ref{relationGchi}) as $X_{\rm min}>-2.77$.

\begin{figure*}
\includegraphics{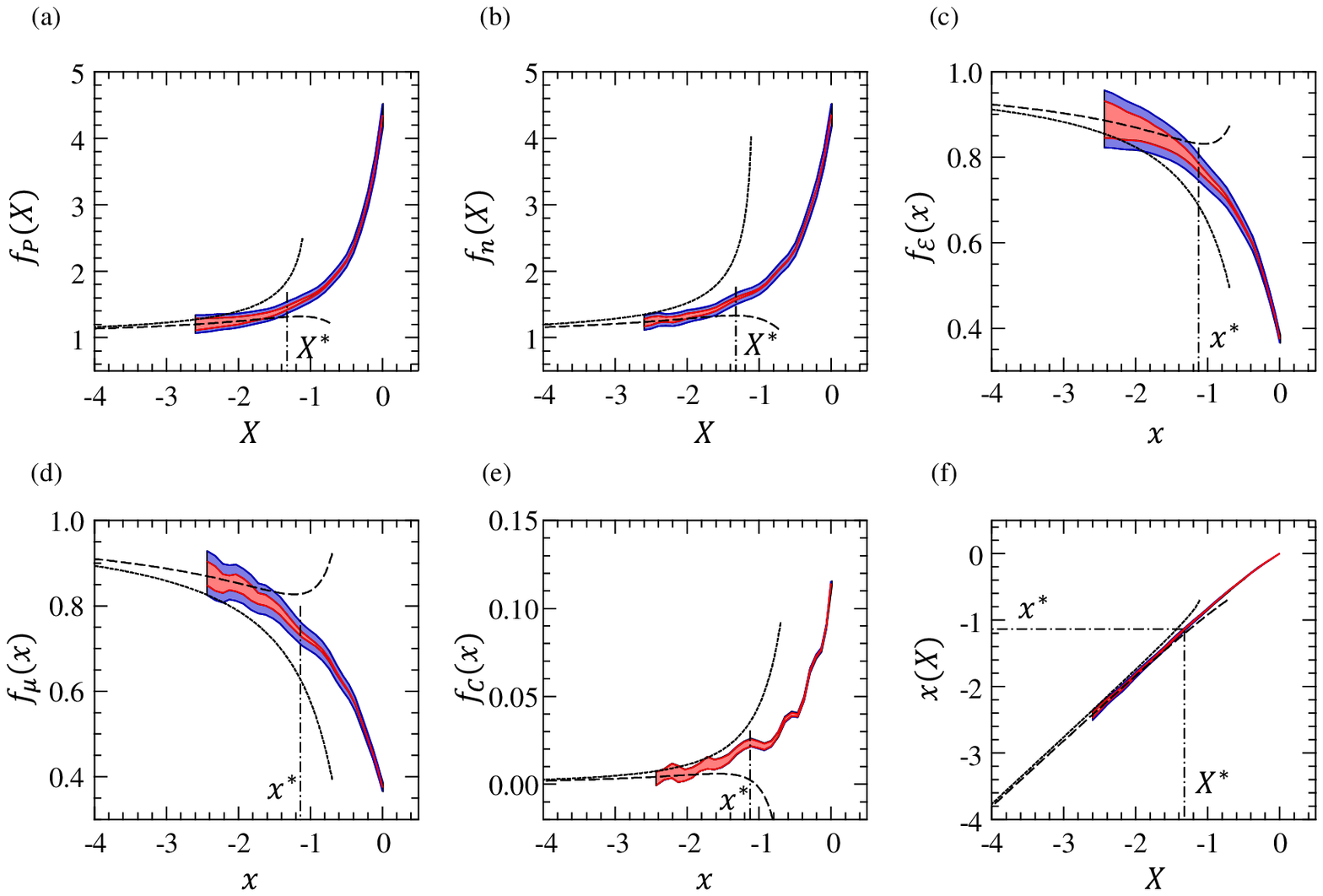}
\caption{\label{fig7}
Dimensionless thermodynamic functions derived from $\mathcal{G}(\chi)$.
(a) Pressure. (b) Number density. (c) Internal energy density. (d) Chemical potential. (e) Contact density. (f) Canonical interaction parameter.
The red colors indicate ranges of statistical errors caused by the uncertainty of the offset value of $\mathcal{G}(\chi)$.
The blue colors indicate additional statistical errors resulting from the uncertainty of the absorption cross-section. The dotted curves and the dashed curves show their asymptotic behavior at the BCS limit up to the first and second order of $1/x$, respectively.
The vertical dash-dot lines indicate interaction parameters $X^*$ and $x^*$ where superfluid transition occurs. The Fermi system is in the superfluid state in the region of $X>X^*$ or $x>x^*$.
}
\end{figure*}

In addition, the experimental data for $f_P$ must satisfy the condition of $f_P>f_P(X_{\rm min})$ from the requirement of $f_P'(X)>0$.
This condition gives an upper bound on $\mathcal{G}(\chi)$ as
\begin{equation}
\mathcal{G}(\chi)<\left( -\frac{15\pi^2}{2f_P(X_{\rm min})}\chi \right)^{2/5}
\label{condition2}
\end{equation}
for $\chi<\chi_{\rm max}$.
In order to use the above inequality, we still need a lower bound on $f_P(X_{\rm min})$.
Here, we invoke the theoretical asymptotic behavior to derive a bound that is better than the obvious bound of $f_P(X_{\rm min})>1$. We confirmed that calculations of $f_P^{\rm Asym}(X)$ up to the first and second orders of $x^{-1}$ in Eq.~(\ref{eq34}) give almost the same values at $X=-5$, which is $f_P(X=-5)=1.1$. By assuming that this value is reliable, we have the bound $f_P(X_{\rm min})>1.1$, because $X_{\rm min}>-5$. Using this bound, the condition of Eq.~(\ref{condition2}) is represented by the dashed red curve in Figure \ref{fig3}a. We confirmed that the entire experimental curve $\mathcal{G}(\chi)$ lies below the red curve if and only if $\mathcal{G}(\chi_{\rm max})<0.14$.

Having established that $0.13<\mathcal{G}(\chi_{\rm max})<0.14$, we chose the offset such that $\mathcal{G}(\chi_{\rm max})=0.135$, which is shown as curve (iii) in Figure \ref{fig3}a. This leads to a systematic error of $|\Delta \mathcal{G}|<0.05$. Note that this constant uncertainty in the offset value becomes negligible toward the unitarity limit, because the value of $\mathcal{G}(\chi)$ around the unitarity limit is approximately four orders of magnitude larger than that around the BCS limit, as shown in Figure \ref{fig3}. In terms of parameter $X$, the relative systematic error $|\Delta \mathcal{G}|/\mathcal{G}(\chi)$ is bounded by $0.05|X|^2$ because $\mathcal{G}(\chi)=|X|^{-2}$ from Eq.~(\ref{relationGchi}).

\subsection{Determination of $f_P(X)$, $f_n(X)$, $f_\mathcal{E} (x)$, $f_\mu(x)$, and $f_\mathcal{C} (x)$}

We converted $\mathcal{G}(\chi)$ to $f_P(X)$ according to Eqs.~(\ref{relationGchi}), and derived the other dimensionless functions from $f_P(X)$ using the thermodynamic relations of Eqs.~(\ref{relationnP} through \ref{relationmuk}).
Figure \ref{fig7} shows the results.
The red regions indicate systematic errors caused by uncertainty $\Delta \mathcal{G}$ for determining the offset value of $\mathcal{G}(\chi)$, which depend on $X$.
The blue regions indicate systematic errors caused by uncertainty of absorption cross-section 
$\sigma_{\rm abs}$, which do not depend on $X$.
In the interaction region of $X>-0.18$ or $x>-0.12$, the former errors become one order smaller than the latter.
Therefore, all of the thermodynamic quantities have been determined by the uncertainty of $\sigma_{\rm abs}$ as  $\frac{\delta f_P}{f_P}=\frac{\delta f_n}{f_n}=\frac{\delta \sigma_{\rm abs}}{\sigma_{\rm abs}}=4\%$ and $\frac{\delta f_\mathcal{E}}{f_\mathcal{E}}=\frac{\delta f_\mu}{f_\mu}=\frac{2}{3}\frac{\delta \sigma_{\rm abs}}{\sigma_{\rm abs}}=3\%$ around the unitarity limit.
Additional systematic errors caused by finite temperature effects in the normal state can be estimated to be around 1\% from the experimental temperature $t_{\rm exp}\lesssim 0.1$ and $t^2$ dependence of thermodynamic quantities.

We found that the superfluid transition point corresponds to $X^*=-1.33$ at $x^*=-1.14$ as shown in Figure \ref{fig7}(f).
The dotted bars at $X^*$ and $x^*$ indicate the superfluid transition points.
Changes in $f_P$, $f_n$, $f_\mathcal{E}$, and $f_\mu$ and a cusp in $f_\mathcal{C}$ around the points can be seen.
Such critical behavior can be explained as appearing the condensation energy at the normal to the superfluid transition point \cite{PhysRevA.88.063614}.
(We have not identified the origin of the other cusps in $f_\mathcal{C} (x)$ around $x=-0.6$ and $x=-0.2$.)
In the present paper, we analyzed data under the assumption of zero temperature.
Therefore, reliable data ranges are the superfluid region at $X>X^*$ and $x>x^*$.
The quantitative evaluation of critical behaviors requires finite-temperature analysis.
This is a subject for future research.

\section{\label{sec:Result}Discussions}

\subsection{\label{Comparison}Comparison with previous experiments and many-body theories}

\subsubsection{Isothermal compressibility: $f_\kappa(x)$}

\begin{figure}
\includegraphics{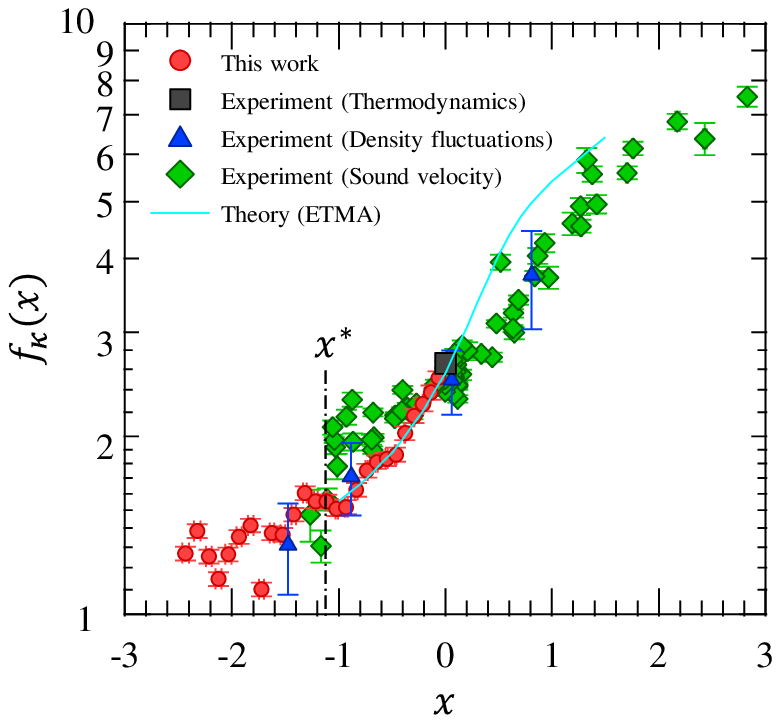}
\caption{\label{fig10}
Experimental and theoretical dimensionless isothermal compressibility.
The vertical axis follows a logarithmic scale.
The red circles indicate the experimental data of the present study.
The green diamonds, blue triangles, and black square indicate experimental values obtained based on the speed of sound \cite{PhysRevLett.98.170401}, density fluctuations \cite{PhysRevLett.106.010402}, and thermodynamic measurements at the unitarity limit \cite{Ku563}, respectively.
The light-blue curve indicates the theoretical calculation of ETMA \cite{PhysRevA.93.013610,TajimaETMA}.
The vertical dash-dot line indicates the superfluid transition point $x^*$ obtained in the present study.
}
\end{figure}

\begin{figure}
\includegraphics{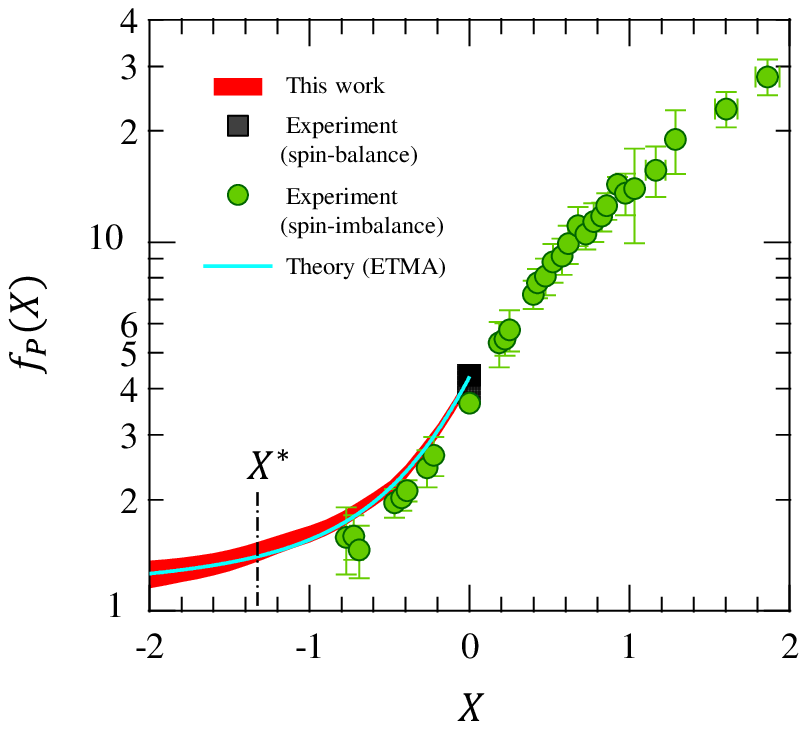}
\caption{\label{fig8}
Experimental and theoretical dimensionless pressure.
The vertical axis follows a logarithmic scale.
The red curve indicates the experimental data of the present study.
The width indicates the total systematic error.
The green circles and the black square indicate the experimental values obtained using spin-imbalanced Fermi gases \cite{Navon729} and spin-balanced unitary Fermi gas \cite{Ku563}, respectively.
The light-blue curve indicates the theoretical calculation of the ETMA \cite{PhysRevA.93.013610,TajimaETMA}.
The vertical dash-dot line indicates the superfluid transition point $X^*$ of the present study.
}
\end{figure}

The obtained dimensionless compressibility is shown by the red circles in Figure \ref{fig10}, along with previously obtained experimental data and a theoretical curve.
The vertical axis follows a logarithmic scale.

At the unitarity limit, $f_\kappa (0)=2.66(7)$ obtained in the preset study is consistent with a previously reported experimental value (2.66(3)), which was determined accurately by measuring the thermodynamics for homogeneous unitary Fermi gases \cite{Ku563}.
The results for two different experiments are shown around the unitarity limit.
One experiment measured density fluctuations \cite{PhysRevLett.106.010402}.
The compressibility was determined based on density fluctuations according to the fluctuation-dissipation theorem. The other experiment measured the speed of sound propagating in the Fermi gas \cite{PhysRevLett.98.170401}.
The speed of sound can be easily converted to isothermal compressibility, as shown in Eq.~(\ref{relationvk}).
While the results of these experiments are not the exact values for a homogeneous system because they were measured for a trapped system, the results of the present study show good qualitative agreement.

The theoretical curve was calculated using an extended $T$-matrix approximation (ETMA) developed in a previous study \cite{PhysRevA.93.013610,TajimaETMA} without using any fitting parameters.
The theory excellently reproduces our experimental results in the superfluid region.

\subsubsection{Pressure: $f_P(X)$}

The determined dimensionless pressure is shown by the red curve in Figure \ref{fig8}, along with previous experimental data and a theoretical curve.
The width of the red curve indicates the overall systematic error.
The vertical axis follows a logarithmic scale.

Our value of $f_P(0)=4.35(17)$ at the unitarity limit is consistent with the experimental value of 4.34(7), which is obtained by $f_P(0)=\xi^{-3/2}$, with $\xi=0.376(4)$ determined using spin-balanced unitary Fermi gases in a previous study \cite{Ku563}.
Our values from the BCS region to the unitarity limit agree qualitatively with the values determined using spin-imbalanced Fermi gases in a previous study \cite{Navon729}, whereas our results are slightly larger.
The ETMA \cite{PhysRevA.93.013610,TajimaETMA} again reproduces our experimental results.

\subsubsection{Internal energy density: $f_\mathcal{E}(x)$}

\begin{figure}
\includegraphics{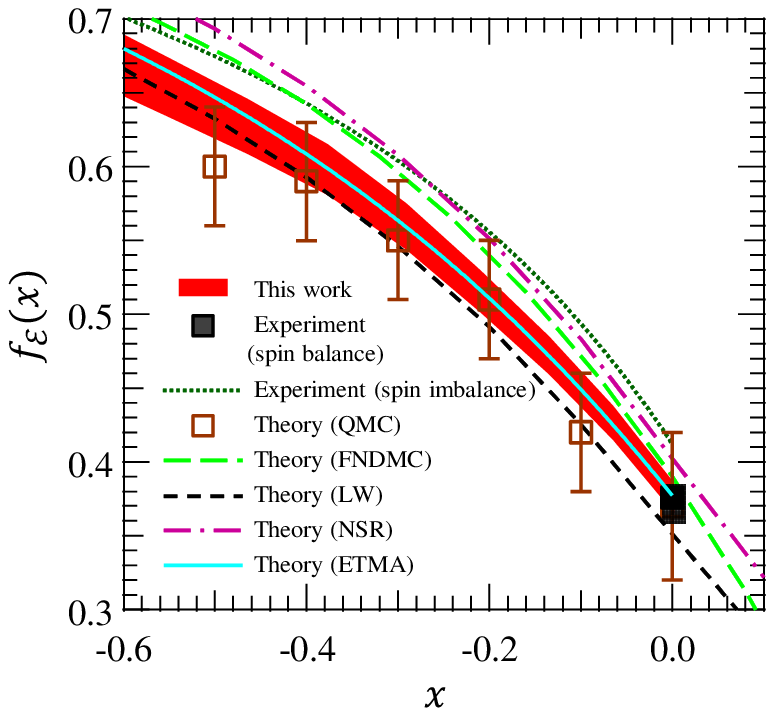}
\caption{\label{fig9}
Experimental and theoretical dimensionless internal energy density.
The red curve indicates the experimental data of the present study in the superfluid state ($x>x^*$).
The width indicates the total systematic error.
The black square indicates the experimental data obtained using spin-balanced unitary Fermi gas \cite{Ku563}, and the dotted green curve is an approximated curve based on pressure measurements of \cite{Navon729}.
The light-blue curve indicates the ETMA \cite{PhysRevA.93.013610,TajimaETMA}. The open brown squares indicate data obtained by the QMC method \cite{PhysRevA.78.023625}. The long dashed green curve was obtained by the FNDMC method \cite{PhysRevA.83.041601}. The dash-dot purple curve was obtained by the NSR theory \cite{0295-5075-74-4-574}. The dashed black curve was obtained by the LW approach \cite{PhysRevA.75.023610}.
}
\end{figure}

The determined dimensionless internal energy density is shown by the red curve in Figure \ref{fig9}, along with previous experimental data and various theoretical curves.
The width of the red curve indicates the overall systematic errors.

At the unitarity limit, our value of $f_\mathcal{E} (0)=0.375(10)$ agrees with the experimental value of 0.376(4) \cite{Ku563}.
In a previous experiment \cite{Navon729}, an approximated curve of $f_P^{\rm Pad\acute{e}}(X)$ was prepared using the Pad\'e approximation. 
An approximated curve $f_\mathcal{E}^{\rm Pad\acute{e}}(x)$ was obtained from $f_P^{\rm Pad\acute{e}}(X)$ using Eqs.~(\ref{relationnP}) and (\ref{relationB}), as shown by the dotted green curve.
We found our data to be inconsistent with this approximated curve $f_\mathcal{E}^{\rm Pad\acute{e}}(x)$.

The results of the ETMA \cite{PhysRevA.93.013610,TajimaETMA} and the quantum Monte Carlo (QMC) method \cite{PhysRevA.78.023625} agree with our result within the error bars.
A fixed-node diffusion Monte Carlo (FNDMC) \cite{PhysRevA.83.041601} method and the Nozi\`eres and Schmitt-Rink (NSR) theory \cite{0295-5075-74-4-574} provide larger values around the unitarity limit. A Luttinger-Ward (LW) approach \cite{PhysRevA.75.023610} shows smaller values than our data around the unitarity limit. 

\subsubsection{Contact density: $f_\mathcal{C}(x)$}

\begin{figure}
\includegraphics{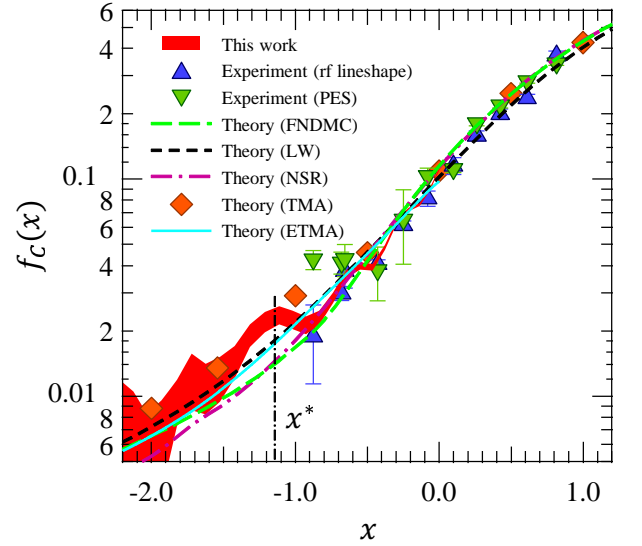}
\caption{\label{fig11}
Experimental and theoretical of dimensionless contact density.
The vertical axis follows a logarithmic scale.
The red curve indicates the experimental data of the present study.
The width indicates the total systematic errors.
The blue triangles and the green inverse triangles indicate experimental values determined by radio frequency spectroscopy (rf line shape) and photon emission spectroscopy (PES) \cite{PhysRevLett.114.075301}. The light-blue curve is the ETMA \cite{PhysRevA.93.013610,TajimaETMA}. The long dashed green curve indicates the results obtained by the FNDMC method \cite{PhysRevA.83.041601}. The dash-dot purple curve indicates the results obtained by the NSR theory \cite{PhysRevLett.105.070402}. The short dashed black curve indicates the LW approach \cite{PhysRevA.80.063612}. The brown diamonds indicate the results obtained by TMA \cite{PhysRevA.82.021605}.
}
\end{figure}

The obtained contact density is shown by the red curve in Figure \ref{fig11}, along with previous experimental data and various theoretical curves. The width of the red curve indicates the overall systematic error. The vertical axis follows a logarithmic scale.
The vertical dash-dot line indicates the superfluid transition point at $x=x^*$.

In the previous experiment \cite{PhysRevLett.114.075301}, contact density was measured using fermions in the normal state at $t=0.18(2) \gtrsim t_c$ by radio frequency spectroscopy (RF line shape) and photon emission spectroscopy (PES). The temperature parameter is approximately three times higher than in our experimental condition.
Nevertheless, the results of the present study agreed quantitatively within the error bars, indicating that the short-range correlation is insensitive to both temperature and whether the many-body state is the normal state or the superfluid state.

We compared our results with the results of the FNDMC method \cite{PhysRevA.83.041601}, the NSR theory \cite{PhysRevLett.105.070402}, the LW approach \cite{PhysRevA.80.063612}, $T$-matrix approximation (TMA) \cite{PhysRevA.82.021605}, and ETMA \cite{PhysRevA.93.013610,TajimaETMA}.
All of these results agreed quantitatively with our data in the BCS region.

\subsection{\label{sec:BCS}Superfluid order parameter}

In several strong-coupling theories, the order parameter ${\it \Delta}$ and chemical potential $\mu$ are calculated for given values of $n$ and $a$ by an equation obtained by combining a gap equation and a number equation. The gap equation is often approximated by the ordinary BCS gap equation \cite{PhysRevA.77.023626, 0295-5075-74-4-574, Levin2010233,PhysRevA.93.013610, TajimaETMA}, whereas the number equation is derived from the single-particle Green's function or the thermodynamic potential, which includes quantum fluctuations. In this case, ${\it \Delta}$ can be uniquely determined by the chemical potential of interacting fermions at $T=0$. If this theoretical approach is adequate, we can calculate ${\it \Delta}$ by solving the gap equation with the chemical potential determined in the present study. Note that it is not obvious that the superfluid order parameter should satisfy the gap equation in the unitary regime. Nonetheless, it is interesting to evaluate ${\it \Delta}$ from thermodynamic quantities using the ordinary BCS gap equation and to compare it with values obtained by other methods.

The BCS gap equation at $T=0$ is given by
\begin{equation}
-\frac{1}{a}=\frac{2}{\pi}\left( \frac{2m{\it \Delta}}{\hbar ^2} \right)^{1/2} I_1\left( \frac{\mu}{\it \Delta} \right),
\label{eqETMA2}
\end{equation}
where the function $I_1(\mu/{\it \Delta})$ is defined in a previous study \cite{Marini1998}.
When we define the dimensionless superfluid gap as
\begin{equation}
f_{{\it \Delta}}=\frac{{\it \Delta}}{\varepsilon_F (n)},\label{eq22}
\end{equation}
the gap equation can be expressed in dimensionless form, relating $x$ and $f_{\it \Delta}$ as
\begin{equation}
1=-\frac{2}{\pi}\frac{\sqrt{f_{\it \Delta}}}{x} I_1\left( \frac{f_\mu(x)}{f_{\it \Delta}} \right)
\label{eq33}.
\end{equation}
Note that ${\it \Delta}$ and $\mu$ determined by the gap equation of Eq.~(\ref{eqETMA2}) have the universal ratio of ${\it \Delta}/\mu=1.16$ at the unitarity limit.

\begin{figure}
\includegraphics{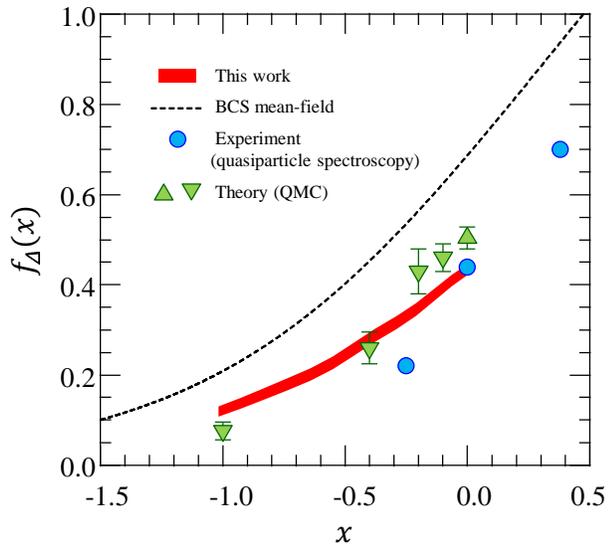}
\caption{\label{fig12}
Dimensionless superfluid gap.
The red curve indicates the gap calculated by substituting our $f_\mu (x)$ into the gap equation in Eq.~(\ref{eq33}).
The dashed black curve was obtained by BCS-MF calculation \cite{Marini1998}.
The green triangle and the green inverse triangles indicate the theoretical values obtained by the QMC method \cite{PhysRevLett.95.060401,PhysRevC.77.032801}.
The blue circles are experimental values obtained by quasiparticle spectroscopy \cite{PhysRevLett.101.140403}.
}
\end{figure}

We substituted $f_\mu(x)$ determined in the superfluid range ($x>x^*$) into Eq.~(\ref{eq33}) and calculated $f_{{\it \Delta}}$ as a function of $x$.
The result is shown by the red curve in Figure \ref{fig12}.
This curve essentially represents the relation among ${\it \Delta}$, $n$, and $a$ under the assumption that the gap equation holds true.
The width of the red curve indicates the overall systematic error.
The obtained values are close to the binding energy of the paired fermions directly observed by a spectroscopic experiment \cite{PhysRevLett.101.140403} as well as that obtained by the quantum Monte Carlo method \cite{PhysRevLett.95.060401,PhysRevC.77.032801}.
Consequently, we found that the binding energy of the paired fermions can be simply estimated by substituting $f_\mu(x)$ into the gap equation.
Note that the above observation does not indicate whether the magnitude of the order parameter obeys the gap equation, because the order parameter can deviate from the binding energy in the unitary regime.
Direct measurement of the order parameter, such as the measurement of the Higgs mode, is desired in this regard \cite{PhysRevA.86.053604}.

\section{\label{sec5}Conclusion}

We determined various thermodynamic quantities in their dimensionless forms from the BCS region to the unitarity limit for homogeneous spin $1/2$ fermions in the superfluid state at the zero-temperature limit.
All of the quantities were determined within systematic errors of 4\% around the unitarity limit based on standard thermodynamic relations, the scale invariant property, and the local density approximation.
In particular, we determined the $f_\mathcal{E}(x)$ for internal energy density directly from experimental data without using an approximated model function.
The dimensionless function represents a universal property shared by various physical systems, and helps to construct the EOS for pure neutron matter, for example.

The thermodynamic quantities obtained here are valuable for studying the strong-coupling properties of a superfluid Fermi gas in the BCS-BEC crossover region.
We evaluated various many-body theories using the obtained thermodynamic quantities.
These theories agree qualitatively with our results, but the measurement results for $f_\mathcal{E}(x)$ allowed us to perform more quantitative comparison among those theories.
The ETMA, which is a strong-coupling theory involving fluctuations in the Cooper channel, provided the closest results to those of the present study. The details of the ETMA can be found in previous studies \cite{PhysRevA.93.013610,TajimaETMA}.

We also found that ${\it \Delta}$ obtained by substituting the chemical potential of interacting fermions into the ordinary BCS gap equation has a value close to the binding energy of the paired fermions.
This is new information regarding the relation between thermodynamic quantities and the binding energy in the BCS region close to the unitarity limit.
Future experiments, such as experiments involving the measurement of the Higgs mode of the order parameter, will reveal the magnitude of the order parameter as well as the relation between the order parameter and the binding energy.

\begin{acknowledgments}

We would like to thank N. Navon, C. Salomon, C. Sanner, W. Ketterle, J. Thomas, T. E. Drake, and D. Jin for providing their experimental data, R. Haussmann, W. Zwerger, H. Hu, G. C. Strinati, A. Gezerlis, S. Gandolfi, and J. Carlson for providing their theoretical calculations, and W. Zwerger for valuable discussions about the critical behavior of contact density around the superfluid transition point. MH would like to thank Y. Aratake for assistance in conducting the experiments. HT and YO would like to thank P. van Wyk, R. Hanai, D. Kagamihara, and D. Inotani for their useful discussions. The present study was supported by a Grant-in-Aid for Scientific Research on Innovative Areas (No. 24105006) and by a Grant-in-Aid for Young Scientists (A) (No. 23684033). HT was supported by a Grant-in-Aid for JSPS Fellows. YO was supported by the KiPAS project of Keio University as well as by Grants-in-Aid for Scientific Research from MEXT and JSPS (No. 15H00840, No. 15K00178, and No. 16K05503).

\end{acknowledgments}

\appendix

\section{\label{sec:Temperature}Superfluid transition point and experimental temperature}

Experimentally, it is impossible to prepare fermions at zero temperature, even if we start from a molecular BEC, and so fermions are always prepared at finite temperature.
In the case of Fermi systems, it is proper to discuss temperature in terms of a temperature parameter defined as $t=k_B T/\varepsilon_F$.
When investigating the properties of interacting fermions in the ground state, the experimental temperature $t_{\rm exp}$ should not only be sufficiently lower than 1 in order to exclude temperature and entropy from the thermodynamics, but should also be lower than the superfluid critical temperature, $t_c$, to take into account the condensation energy \cite{PhysRevA.88.063614}.
Therefore, $t_{\rm exp}<t_c<1$ should be satisfied in order to investigate the ground state.
Since $t_c$ decreases monotonically to zero from the unitarity limit to the BCS limit as a function of the interaction parameter $x$, it is inevitable that $t_c (x)$ intersects $t_{\rm exp}$ somewhere between the two interaction limits.
Here, we define the intersection as $t_{\rm exp}=t_c (x^* )$.

We determined the superfluid transition point $x^*$ by measuring the condensate fraction (CF) of paired fermions. The value of $x^*$ can be determined even if we know neither the experimental temperature $t_{\rm exp}$ nor the function for the critical temperature $t_c (x)$.
Since the local density has a peak value at the bottom of the trap, the ratio $t_{\rm exp}/t_c$ takes the minimum value there.
Therefore, CF takes finite values when fermions satisfy $t_{\rm exp}/t_c<1$ at the bottom, and CF becomes zero at $t_{\rm exp}=t_c (x^* )$.

We carried out the measurement of CF as follows.
We prepared ultracold fermions at various Feshbach magnetic fields using the same experimental procedure as in the present study.
Instead of measuring the {\it in-situ} column density of the fermions, we measured the center-of-mass (CM) momentum distribution of the paired fermions \cite{HoriSuperfluid}.
We turned off the ODT and the magnetic field simultaneously within 10~$\mu$s.
At that time, the paired fermions were converted to molecules because the Fermi system was changed to the BEC regime.
These molecules expanded with the original CM momentum distribution of the paired fermions during time-of-flight (TOF).
We measured the momentum distribution after a TOF of 11 ms by absorption imaging. 

\begin{figure}[!tb]
\includegraphics{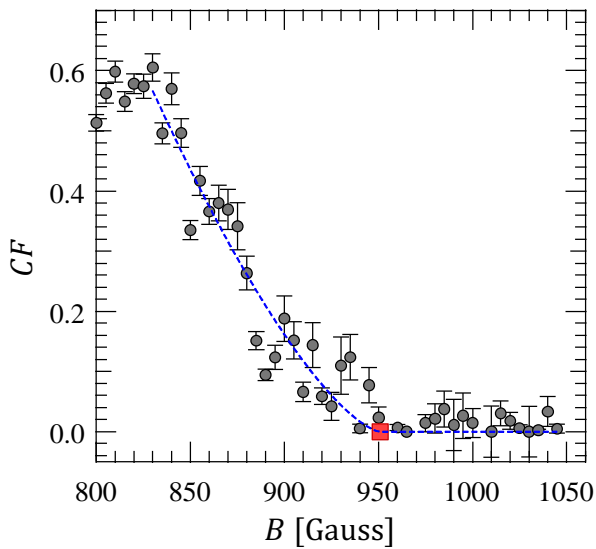}
\caption{\label{fig5}
Condensate fraction of paired fermions evaluated at various Feshbach magnetic fields.
The circles with error bars indicate experimental data.
The dashed curve shows the fitting result.
The red square corresponds to the point at which the condensate fraction disappears.
}
\end{figure}

Figure \ref{fig5} shows the CF evaluated by fitting a bimodal function to the measured momentum distribution.
In order to estimate the magnetic field $B^*$, at which CF becomes zero, we fit an empirical function of $CF=a(B^*-B)^b \cdot {\rm \Theta}(B^*-B)$ to the data with fitting parameters $B^*$, $a$, and $b$.
The fitting result yielded $B^*=$951 Gauss.
We then evaluated the value of $x^*$ by $x^*=\frac{1}{k_F(n)a(B^*)}$ with the peak density and the scattering length at the magnetic field.
In this way, the superfluid transition point was determined to be $x^*$=$-$1.14.
In the region of $x>x^*$, fermions are in the superfluid state.

Next, we estimated the experimental temperature $t_{\rm exp}$.
At the unitarity limit (832.18~Gauss), we observed a condensate fraction of 0.6, as shown in Figure \ref{fig5}.
This suggests that $t_{\rm exp} < 0.1$ at $x=0$, according to a previous experimental result \cite{Ku563}.
At $x^*=-1.14$, the critical temperature was calculated to be $t_c (x^* )=0.06$ \cite{PhysRevA.75.023610}.
Since $t_{\rm exp}=t_c (x^* )$, we estimated ${t_{\rm exp}\sim 0.06}$ at $x=-1.14$.
At $x<-1.14$, it is difficult to estimate the temperature parameter from our experimental data.
According to a theoretical work of Ref.~\cite{PhysRevLett.95.260405}, the temperature parameter at the trap center decreases during an adiabatic change from the BEC region to the BCS region along the isentropic curve.
Consequently, we estimated the temperature parameter to be $0.06\lesssim t_{\rm exp}\lesssim 0.1$ from around the BCS region to the unitarity limit.

\bibliography{GroundStateRef}

\end{document}